# COUPLING OF MAGNETIC ORDER, FERROELECTRICITY, AND LATTICE STRAIN IN MULTIFERROIC RARE EARTH MANGANITES

B. Lorenz, C. R. dela Cruz, F. Yen, Y. Q. Wang, Y. Y.Sun, C. W. Chu[*]

TCSUH and Department of Physics, University of Houston
3201 Cullen Blvd.
Houston, TX 77204-5002

[*] also at
LBNL Berkeley
1 Cyclotron Road
Berkeley, CA 94720
and
Hong Kong University of Science and Technology
Hong Kong, China

ABSTRACT
    Multiferroic rare earth manganites attracted recent attention because of the coexistence of different types of magnetic and ferroelectric orders resulting in complex phase diagrams and a wealth of physical phenomena. The coupling and mutual interference of the different orders and the large magnetoelectric effect observed in several compounds are of fundamental interest and bear the potential for future applications in which the dielectric (magnetic) properties can be modified by the onset of a magnetic (dielectric) transition or the application of a magnetic (electric) field. The physical mechanisms of the magnetoelectric effect and the origin of ferroelectric order at magnetic transitions have yet to be explored. We discuss multiferroic phenomena in the hexagonal $HoMnO_3$ and show that the strong magneto-dielectric coupling is intimately related to the lattice strain induced by unusually large spin-phonon correlations.

INTRODUCTION
    The coexistence and mutual interference of different types of long-range orders, such as magnetic, elastic, and ferroelectric (FE), have long inspired researchers because of their fundamental interest and their significance for potential applications. The coupling between ferroelectricity and magnetism recently observed in rare earth (R) manganites such as hexagonal $RMnO_3$[1-9], orthorhombic $RMnO_3$[10-12], $RMn_2O_5$[13-27], and other compounds[28-35] has attracted increasing attention because of the prospect of controlling the dielectric (magnetic) properties of these materials by an external magnetic (electric) field. These effects are particularly strong in materials with coexisting magnetic and dielectric orders (multiferroic materials) and close to magnetic or ferroelectric phase transitions. The number of compounds exhibiting multiferroic properties is limited and possible reasons for this are discussed in Ref. [36]. The coexistence of ferroelectric and magnetic orders requires the simultaneous breaking of the spatial inversion and time reversal symmetries. Although the original definition of multiferroicity demands the coexistence of ferroelectric, ferromagnetic and/or ferroelastic orders it was later extended to include compounds exhibiting antiferromagnetic (AFM) order and ferroelectricity. Most of the manganites listed above fall into this category with either ferroelectricity arising at $T_c$ well above

the AFM Néel temperature ($T_N$), for example all the hexagonal RMnO$_3$ (R=Ho to Lu and Y), or FE being stabilized below or close to an AFM transition, as observed in some orthorhombic RMnO$_3$ (R=Tb and Dy) and in most of the RMn$_2$O$_5$. Other rare earth manganites, although not ferroelectric, exhibit pronounced anomalies of the dielectric constant at the magnetic phase transitions indicating a strong coupling of the AFM order with the lattice degrees of freedom[11,12].

Another common and interesting feature of multiferroic rare earth manganites is the magnetic frustration that leads to a non-collinear spin alignment in the AFM phase (e.g. in hexagonal RMnO$_3$) and to incommensurate magnetic order with commensurate lock-in transitions at lower temperature (RMn$_2$O$_5$). The spin frustration originates either from geometric constraints as in the case of the hexagonal P6$_3$cm structure[3,37] (three neighboring Mn$^{3+}$ spins form an equilateral triangle resulting in the frustration of their AFM exchange interactions) or from competing Mn$^{3+}$ exchange interactions between nearest and further distant neighbors in the orthorhombic structures[15,38]. In addition to these complicated interactions among the Mn ions most of the R$^{3+}$ ions carry their own magnetic moment (e.g. Ho$^{3+}$: J=8, but not Lu$^{3+}$ or Y$^{3+}$) that is usually non-collinear with the Mn spins. For example, in hexagonal RMnO$_3$ the easy axis anisotropy of the rare earth ions aligns the R moments along the c-axis whereas the easy plane anisotropy of the Mn spins constrains the moments strictly into the basal a-b plane. At low temperatures the exchange coupling between both magnetic subsystems becomes stronger resulting in additional distinct changes of the magnetic structure. The antisymmetric Dzyaloshinski-Moriya exchange interaction between f-moments and d-spins perpendicular to each other, for example, tends to stabilize a weak ferromagnetism. The pseudo-dipolar magnetic interaction between perpendicular magnetic moments causes one magnetic subsystem to reorient at a critical temperature and to become aligned with the other magnetic system[39].

Similar effects have been observed in some RMnO$_3$. In orthorhombic HoMnO$_3$ (this structure is metastable and can be synthesized under high-pressure conditions) the Ho$^{3+}$ moments tilt away from their original alignment with the c-axis towards the a-axis (the principal direction of the Mn$^{3+}$ spins) below the AFM ordering temperature of the Mn spins[40] and large magneto-dielectric coupling was observed in this phase[12]. The hexagonal form of HoMnO$_3$ (this is the thermodynamically stable structure) exhibits a series of phase transitions, starting with a FE transition well above room temperature followed by a frustrated non-collinear AFM order of the Mn spins at 76 K. At lower temperature the coupling of the Mn spins to the Ho moments results in an in-plane rotation of the Mn spins and the onset of the AFM order of the Ho. Two more magnetic phase transitions have been reported at 33 K and 5.2 K, respectively[3]. The complex system of FE polarization, Mn spins, and Ho magnetic moments, their correlations and mutual interactions constitute an exciting topic for further investigation and reveal a wealth of interesting physical phenomena resulting in a complex magnetic and dielectric phase diagram. The following sections will be devoted to this interesting compound, its multiferroic properties, and the coupling of magnetic and ferroelectric orders via lattice strain and the magnetoelastic effect.

THE HEXAGONAL HoMnO$_3$: STRUCTURAL ASPECTS AND MAGNETIC ORDER

The crystalline structure of the hexagonal rare earth manganites at ambient temperature is hexagonal, space group P6$_3$cm. It develops from a high temperature (centrosymmetric and paraelectric) phase with P6$_3$/mmc symmetry into an antiferroelectric phase (P6$_3$cm) and eventually into the ferroelectric phase through the tilting of the MnO$_5$ bitetrahedra and a corrugation of the R ion layers[41]. The P6$_3$cm structure of HoMnO$_3$ is shown in Fig. 1. The Mn ions form a triangular lattice in the a-b plane and subsequent planes stacked along the c-axis are offset by a/3 along the a-axis. The ferroelectricity arises mainly from an off-center displacement of the $Ho^{3+}$ ions with an effective polarization along the c-axis.

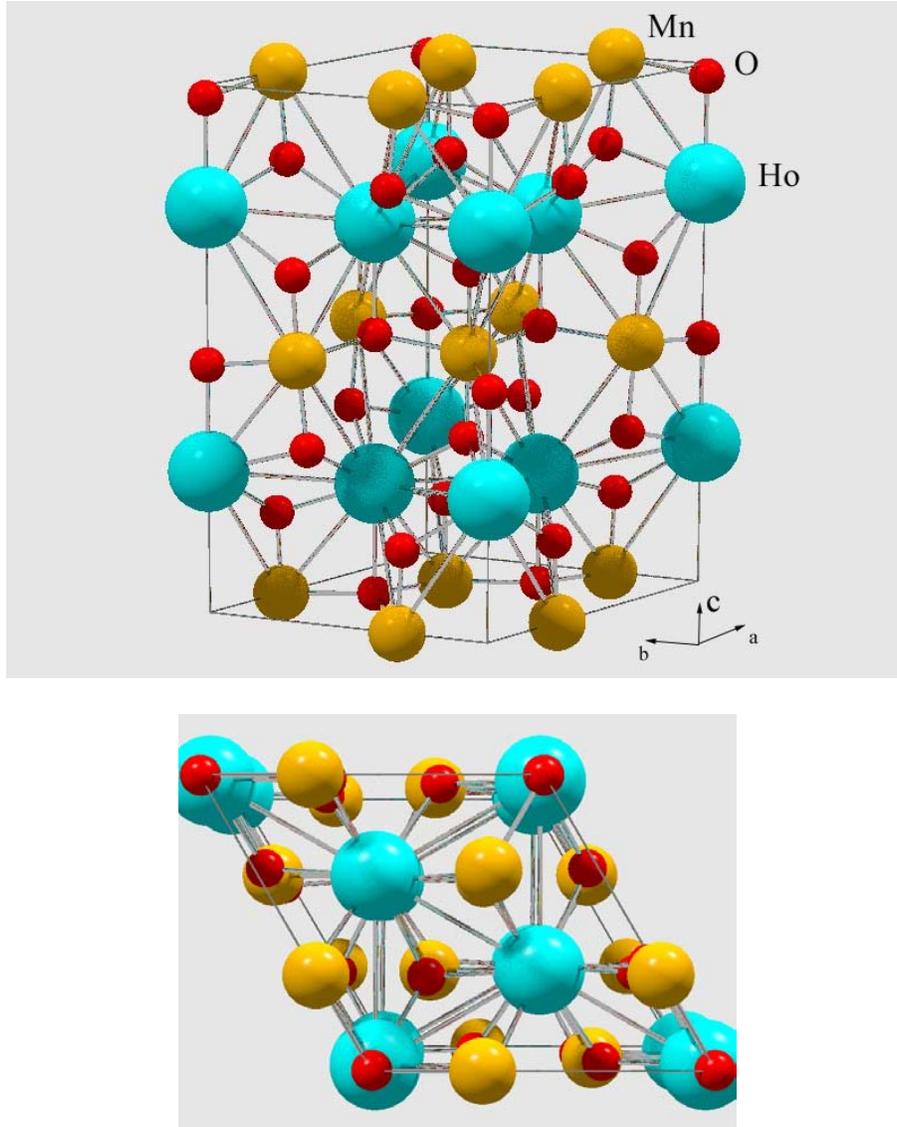

Fig. 1: $P6_3cm$ structure of $HoMnO_3$. The lower figure reveals the triangular sublattice of Mn and Ho in the hexagonal a-b plane.

The Mn spins interact via strong in-plane AFM superexchange interactions whereas the strength of the inter-plane exchange is two orders of magnitude lower[42]. This is an excellent example of a nearly two dimensional Heisenberg type magnet on a triangular lattice that is well known for its spin frustration. Accordingly, the Mn spins undergoing long range order at $T_N$ form a non collinear spin structure characterized by an angle of 120° between neighboring spins. Fiebig et al.[43] have discussed four basic spin configurations that are compatible with the crystalline symmetry. Neutron scattering[44,45,46] and optical second harmonic generation experiments[43] have mainly contributed to resolve the magnetic structures below the Néel temperature $T_N$=76 K of HoMnO$_3$. The results are summarized as follows:

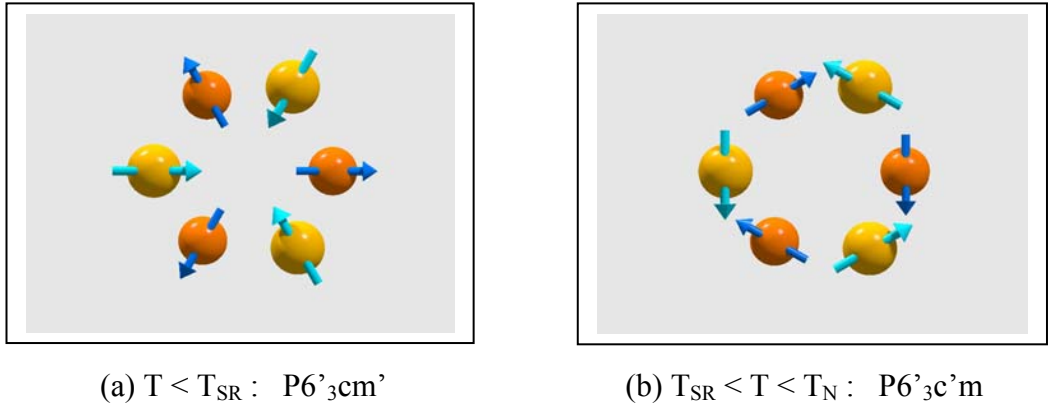

(a) $T < T_{SR}$ :  P6'$_3$cm'    (b) $T_{SR} < T < T_N$ :  P6'$_3$c'm

Fig. 2: Magnetic structure of the Mn spins above and below the spin reorientation transition. Two subsequent layers are shown, the top layer displaying the brighter ions and spins.

Below $T_N$ the Mn spins order in the non collinear way perpendicular to the hexagonal a and b axes shown in Fig. 2b. At $T_{SR}$=33 K all spins rotate by 90° and form the magnetic structure of Fig. 2a. Simultaneously, the Ho moments become magnetically polarized and a small AFM sublattice magnetization was shown to arise and to increase with decreasing temperature[44,45]. The c-axis magnetic susceptibility decreases abruptly by a small amount at $T_{SR}$ indicating the onset of the AFM Ho moment order[3]. At lower temperature, close to 5 K, another spin reorientation transition takes place to a low temperature phase with P6$_3$cm magnetic symmetry accompanied by a strong increase of the Ho AFM moment[44,45,46]. Additional magnetic structures have been proposed to be stable only under external magnetic fields[43].

The complex phase relations observed so far in HoMnO$_3$ are due to the coupling of the Mn spins, the Ho moments, and the ferroelectric polarization. In particular the coupling of the magnetic order with the FE polarization is of fundamental interest and should be detectable in measurements of the dielectric properties. Although the direct coupling between the in-plane magnetic order and the c-axis polarization is not allowed by symmetry[47] in the magnetic structures of Fig. 2 secondary effects mediated by lattice strain and the magnetoelastic effect (spin-lattice interaction) may result in dielectric anomalies at the magnetic phase transitions of HoMnO$_3$.

DIELECTRIC ANOMALIES IN HoMnO$_3$: THE COMPLEX PHASE DIAGRAM

In the search for magneto-dielectric effects in HoMnO$_3$ we have measured the dielectric constant of single crystals grown from the flux. Small platelets between 50 μm and 100 μm thick have been selected for capacitance measurements. Gold contacts evaporated to the two parallel faces formed the capacitor and the AH2500A high resolution capacitance bridge (Andeen-Hagerling) was employed for data acquisition. The sample was mounted to a home-made low-temperature probe that was adapted to Quantum Design's Physical Property Measurement System for the control of temperature between 1.8 K and 300 K and magnetic fields up to 70 kOe. The dielectric constant, ε(T), at zero magnetic field is shown in Fig. 3. Three anomalies are clearly visible. At the Néel transition temperature ε shows a distinct decrease due to the onset of long-range AFM order of the Mn spins. This decrease has also been reported for other hexagonal RMnO$_3$ and it is characteristic for the transition into the AFM ordered state[1,4,5,6]. At the low-temperature end the transition into the P6$_3$cm magnetic structure is marked by a sharp increase of ε(T) at T$_2$ = 5.2 K. The most notable anomaly of ε(T) is the sharp peak at T$_{SR}$ = 32.8 K (Fig. 3). The peak width of only 0.5 K proves the sharpness of the spin reorientation transition and the uniformity and high quality of the crystals used for the present investigation. It is remarkable that ε(T) exhibits well defined and sharp anomalies at all three magnetic phase changes at H=0 whereas other physical quantities such as the dc magnetization exhibit only minor changes that are sometimes impossible or very difficult to detect[2,3,44]. Dielectric measurements can, therefore, be employed as an extremely sensitive probe to detect phase transitions and to investigate the magnetic phase diagram of HoMnO$_3$. It also proves the intimate correlation of ferroelectric and magnetic orders the possible origin of which will be discussed below.

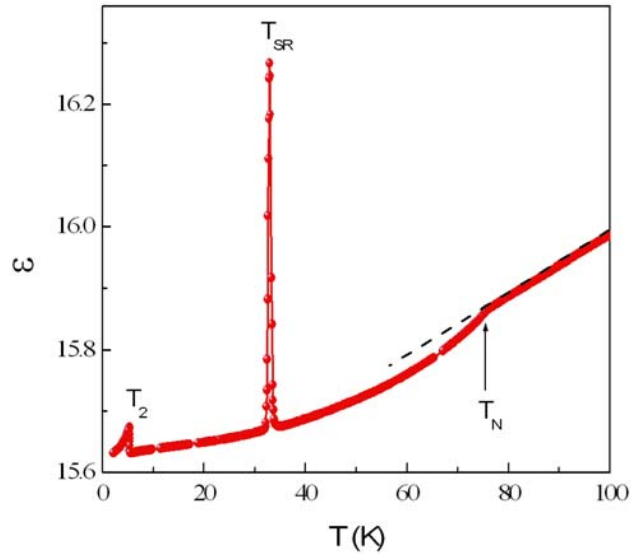

Fig. 3: Dielectric constant of HoMnO$_3$ at zero magnetic field.

The sensitivity of ε(T) with respect to subtle changes of the magnetic order leads us to investigate the phase diagram for magnetic fields H oriented along the c-axis through ε(T, H). Fig. 4 shows the field effect on ε(T) up to 70 kOe. The narrow peak at T$_{SR}$ shifts to lower T with increasing H and develops a plateau-like structure clearly seen above 20 kOe (Fig. 4a). The low-temperature peak at T$_2$ shifts to higher T with H and develops a similar plateau that merges with the high-T plateau at about 33 kOe. Above 40 kOe all anomalies of ε(T) disappeared except a small but sharp drop at about T$_4$ = 4 K (Fig. 4e). At low temperatures ε(T) exhibits additional anomalies (Fig. 4b to e) indicating an unprecedented phase complexity in this region of the phase diagram.

The plateau-like enhancement of ε at fields below 40 kOe reveals the existence of an intermediate phase (INT) between the P6'$_3$c'm (HT1 phase) and P6'$_3$cm' (HT2 phase) magnetic structures as shown in the phase diagram of Fig. 5a. This INT phase covers a larger area in the phase diagram at higher magnetic fields and its phase boundaries, T$_1$(H) and T$_2$(H), are well defined by the sharp changes of ε(T). It is remarkable that ε(T, H) in the INT phase is a well defined function of T, its H-dependence is small, and it follows the dashed line in Fig. 4a.

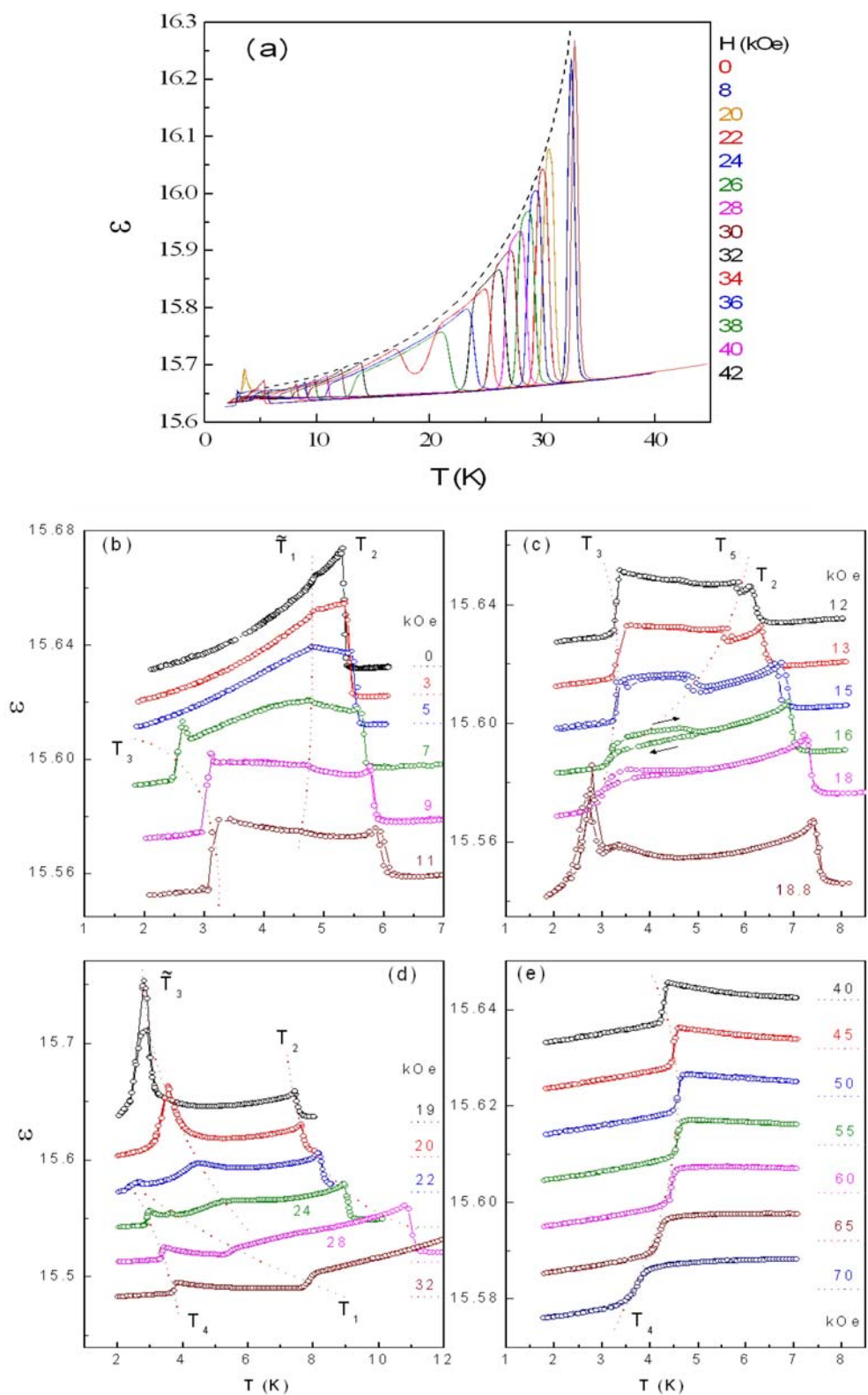

Fig. 4: Low-temperature dielectric constant at magnetic fields up to 70 kOe. The dielectric anomalies at different phase boundaries are indicated by dotted lines denoted by $T_1$ to $T_5$.

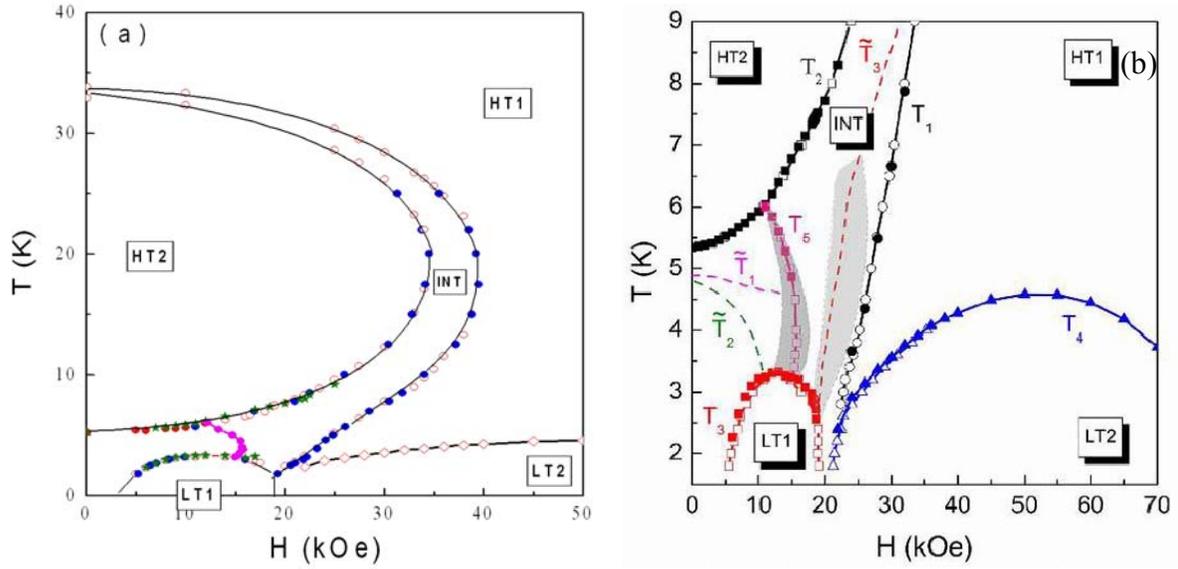

Fig. 5: Phase diagram of HoMnO$_3$. The shaded areas indicate regions where thermal and magnetic field hysteresis was observed. The phase boundaries are labeled by $T_1$ to $T_5$.

The low-temperature (T < 8 K) section of the phase diagram is very complex and the different phase boundaries are derived from the dielectric data shown in Fig. 4b-e. The sharp increase of ε(T) at $T_2$ is shifted to higher T with H. Above 5 kOe there arises another sharp decrease of ε at $T_3$ (Fig. 4b). $T_3$(H) increases with H, passes through a maximum at 12 kOe and decreases again, defining a dome shaped low-temperature phase (LT1). At 12 kOe a small but distinct step develops at $T_2$ and is shifted to lower T with increasing field until it merges with the $T_3$-anomaly (Fig. 4c). The characteristic temperature, denoted by $T_5$, exhibits thermal and magnetic field hysteresis as indicated by the shaded area in Fig. 5b. With further increasing H another step-like change of ε(T) is detected at the lowest temperatures for H > 20 kOe (Fig. 4d). The critical temperature of this anomaly, $T_4$(H), can be traced to the maximum field of 70 kOe (Fig. 4d-e). $T_4$ exhibits a maximum of 4.6 K at 52 kOe and decreases again to higher fields defining a second dome-shaped phase, LT2. Additional features of ε(T, H) are indicated by dashed lines in Fig. 5b and labeled with $\tilde{T}_1$ to $\tilde{T}_3$, however, they may not represent additional phase boundaries but rather indicate subtle changes or smooth crossover phenomena of the magnetic structure of HoMnO$_3$. $\tilde{T}_1$ is defined by a sharp slope change of ε(T) as indicated in Fig. 4b. $\tilde{T}_2$ defines a broad maximum of the isothermal ε(H) as shown in Fig. 6. A sharp and high peak of ε(T, H) arises close to 3 K and 20 kOe and its position is denoted by $\tilde{T}_3$. This peak broadens and shifts to higher T with increasing field as shown in Figs. 4c-d and 6.

The multitude of dielectric anomalies and phase transitions at low T is clearly seen in the H-dependence of the isothermal ε(H) shown in Fig. 6 close to 4.5 K. The anomalies discussed above are marked by vertical arrows attached to the 4.4 K data and labeled by $\tilde{T}_2$ etc. Notably the two maxima at $\tilde{T}_2$ and $\tilde{T}_3$, but also all other phase boundaries in this temperature range ($T_5$, $T_1$, and $T_4$) are unambiguously derived from ε(H) and they coincide with the data extracted from the ε(T) anomalies. At 4.4 K the transition into the LT2 phase at 42 kOe has reentrant character and at 62 kOe the $T_4$ phase boundary is crossed a second time reentering the HT1 phase. This behavior is in perfect agreement with the maximum of $T_4$(H) as displayed in Fig. 5b.

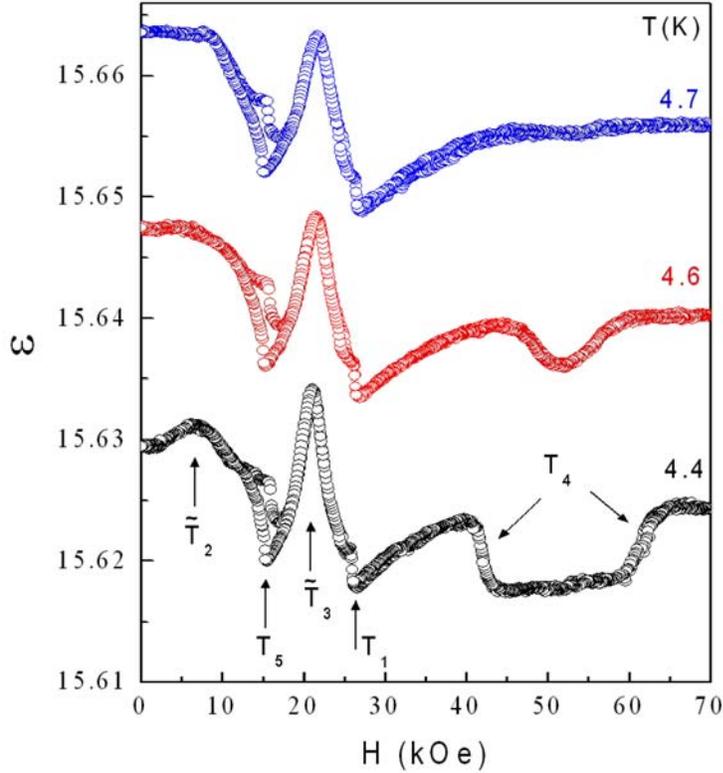

Fig. 6: Low-temperature isothermal dielectric constant, $\varepsilon(H)$. Different anomalies of $\varepsilon$ are marked by arrows and labeled according to the notations of Fig. 5b. Note the hysteresis around $T_5$. The curves are offset by a constant for better clarity.

The phase diagram of Fig. 5 is far more complex than previously assumed[43] and it also accommodates all low-temperature anomalies detected in the magnetic peak intensities of recent neutron scattering experiments[46]. The phase boundaries derived from the dielectric data are in perfect agreement with our previous conclusions based on magnetic data and heat capacity measurements[3]. It is particularly interesting that our results clearly distinguish at least 6 different phases below the Néel temperature in contrast to the 4 magnetic structures (P6'$_3$c'm, P6'$_3$cm', P6$_3$cm, and P6$_3$c'm')[43] that are compatible with the crystalline space group P6$_3$cm. This leads to the conclusion that some of the magnetic structures below $T_N$ have to be of lower symmetry, for example P6'$_3$ (the Mn$^{3+}$ spins form an angle between 0 and 90º with the hexagonal a-axis in this symmetry). The involvement of the Ho$^{3+}$ moment order has to be taken into account at low T. Magnetic measurements[2,3] and neutron scattering data[44,45] provided evidence for the onset of AFM Ho moment order at $T_{SR}$ with a major increase of the Ho moment at $T_2$. At $H > 5$ kOe a sharp increase of the c-axis magnetization and a narrow heat capacity peak at $T_3$ provide convincing evidence for further dramatic changes of the magnetic orders[3].

The magnetic symmetry of the various phases cannot be derived from dielectric measurements and the following discussion has to rely on published results from neutron scattering[44,45,46] and optical experiments[43]. According to these measurements the magnetic structures of the HT1, HT2, and the low-temperature phase that is bound by $T_2(H)$-$T_5(H)$-$T_3(H)$ have been assigned the P6'$_3$c'm, P6'$_3$cm', and P6$_3$cm magnetic symmetries, respectively. The P6$_3$c'm' magnetic space group was proposed for the LT2 phase stable above 20 kOe[43]. This leaves the magnetic structure in the INT and LT1 phases open for discussion. It appears conceivable that the magnetic order in the INT phase, that is sandwiched between the P6'$_3$c'm and P6'$_3$cm' structures, is described by the P6'$_3$ magnetic symmetry. In this symmetry the angle Φ between the $Mn^{3+}$ spin and the a-axis adopts an intermediate value between 0 (corresponding to P6'$_3$cm') and 90º (P6'$_3$c'm). Various quantities such as the dielectric constant[2] or the ac magnetic susceptibility[3] are well-defined functions of temperature and magnetic field in this phase. We, therefore, propose that the angle Φ changes continuously as a function of T and H in the INT phase and that the physical properties are mainly determined by Φ. More detailed investigations of the magnetic structure are needed to verify our proposal. The magnetic order in the LT1 phase is yet unexplored. The transition across $T_3(H)$ is a strong first order phase transition and it is accompanied by large anomalies of the magnetization and the heat capacity[3]. Based on previous data we proposed that a major change of the Ho moment order takes place at $T_3$ but details have yet to be investigated.

The phase diagram shown in Fig. 5 exhibits several multicritical points wherever different phase boundaries join together. Two tricritical points are located at the intersection of $T_5(H)$ with $T_2(H)$ and $T_3(H)$, respectively. Near 20 kOe and below 3 K several phase boundaries come very close and there is the possibility that three transition lines, $T_1$, $T_3$, and $T_4$, merge at lower T, possibly at T=0. This would result in a tetracritical point close to or at zero temperature, an unusual phenomenon of considerable interest. Detailed investigations at ultra low temperature are therefore desired to elucidate the most interesting part of the phase diagram.

THE COUPLING BETWEEN MAGNETIC ORDER AND DIELECTRIC PROPERTIES: EVIDENCE FOR STRONG SPIN-LATTICE INTERACTION

The origin of the coupling of magnetic and ferroelectric orders in multiferroic manganites is currently a matter of discussion. The linear magneto-electric effect is frequently not allowed by symmetry, such as in the case of $HoMnO_3$ in both high-T phases (P6'$_3$c'm and P6'$_3$cm')[47]. Secondary effects mediated by the magnetoelastic distortion of the lattice have consequently been proposed to explain the dielectric anomalies at the AFM transitions[2]. The search for tiny structural distortions using x-ray or neutron scattering is often unsuccessful because of the limited relative resolution of the diffraction techniques. Thermal expansion measurements employing high precision capacitance dilatometers, however, achieve orders of magnitude higher resolution. The thermal expansivities of $HoMnO_3$ along the principal crystallographic directions have been measured in order to detect the smallest magnetoelastic stain of the lattice that is correlated with the observed dielectric anomalies. The results are shown in Fig. 7 for the a- and c-axis thermal expansivities. At the Néel temperature both lattice parameters show strong anomalies, *c* is expanding whereas the in-plane parameters shrink with the establishment of AFM long range order. The expansivities (left inset of Fig. 7) show distinct lambda-type anomalies with opposite signs as expected for a second order phase transition. A similar lambda-shape of the specific heat peak at $T_N$ was observed recently[3]. The most remarkable feature is the negative expansivity of the *c*-axis at all temperatures below room temperature (the *c*-axis expands upon cooling). This unusual property as well as the large anomalies at $T_N$ indicate the existence of very strong spin-spin correlations and spin-lattice coupling at temperatures far above $T_N$. We propose the following qualitative explanation for the observed effects:

in-plane spin correlations couple to the lattice via the magnetoelastic effect and cause a shrinkage of the *a*- and *b*-axis since the magnetic system gains exchange energy when the distance between neighboring spins is reduced. With decreasing temperature the spin-spin correlations increase and so do the magnetostrictive effects resulting in a magnetic contribution to the in-plane thermal expansion and an enhancement of the expansivity. The *c*-axis is affected indirectly via the elastic properties of the lattice. The enhanced contraction of the *a-b* plane causes the *c*-axis to expand upon cooling. This effect obviously is stronger than the common positive contribution to the *c*-axis expansivity that is due to the lattice anharmonicities. The volume expansivity is positive over the whole temperature range. The magnetostrictive effects discussed above are particularly strong at $T_N$ where long range order sets in. The strong anomalies and the opposite sign of the thermal expansivity peaks provide convincing support to our conclusions. Similar thermal expansion anomalies including the negative *c*-axis expansivity have also been observed in $YMnO_3$ and it is possibly a general feature of hexagonal rare earth manganites[48].

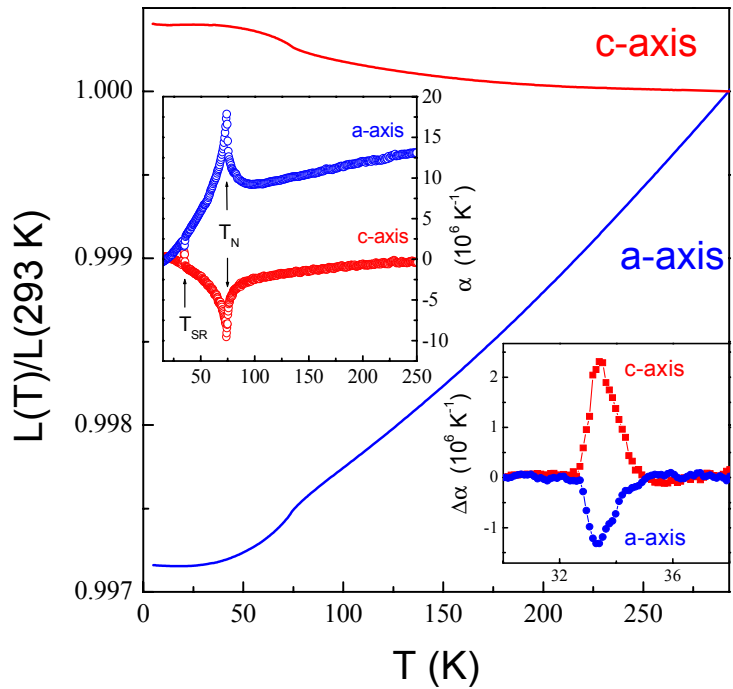

Fig. 7: Thermal expansion of the a- and c-axis of $HoMnO_3$. Left inset: Thermal expansivities below room temperature. Right inset: Expansivities neat $T_{SR}$ (the lattice contribution was subtracted).

At the spin reorientation transition the *a*- and *c*-axis expansivities show sharp peaks of opposite sign indicating that strong spin-lattice couplings still exists at $T_{SR}$ (right inset, Fig. 7). In contrast to the Néel transition, the *c*-axis shrinks whereas the *a*-axis expands upon cooling through $T_{SR}$. This can be understood as an effect of the AFM $Ho^{3+}$ moment order that happens at $T_{SR}$. The exchange correlations of the Ho moments along *c* and the onset of long range order causes the c-axis to shrink at $T_{SR}$. It is interesting to note that the volume below $T_{SR}$ is larger than above resulting in a negative pressure coefficient of $T_{SR}$ as confirmed recently[49]. The narrow peaks of the thermal expansivities as well as of the specific heat observed at $T_{SR}$ evidence the first order nature of the spin reorientation transition. In a detailed analysis of magnetization, heat capacity, and thermal expansivities as well as the pressure and magnetic field dependence of $T_{SR}$ we could show that the entropy balance as required by an extension of the Clausius-Clapeyron equation is fulfilled proving the first order nature of the transition[49].

The physical origin of the coupling between magnetic order and dielectric properties (such as the decrease of $\varepsilon(T)$ below $T_N$) is still a matter of discussion. The linear magneto-electric effect is not allowed by symmetry in the P6'$_3$c'm and P6'$_3$cm' phases. The strong spin-lattice interaction, however, provides evidence for an indirect coupling via lattice strain. This problem has been approached theoretically by modeling the magnetic interactions by an anisotropic Heisenberg model, the ferroelectric order by a double well potential, and the spin-lattice interaction by a term coupling the spin-spin correlation with the ferroelectric displacement coordinate[50]. Within a mean field approximation the model was evaluated and qualitative agreement with the observed dielectric response was achieved for an appropriate choice of the model parameters. According to these results the dielectric constant, $\varepsilon(T)$, is proportional to the inverse square of the ferroelectric displacement coordinate[50]. Since the FE polarization and the displacement coordinate are directed along the $c$-axis it appears justified to search for a correlation of $\varepsilon(T)$ and the inverse square of $c(T)$. This correlation, in fact, exists as convincingly demonstrated in Fig. 8. The scaling of $\varepsilon(T)$ with $1/c^2$ provides the clearest evidence for the importance of the spin-lattice coupling and the resulting strain in understanding the correlation between magnetic order and dielectric properties in hexagonal $R$MnO$_3$.

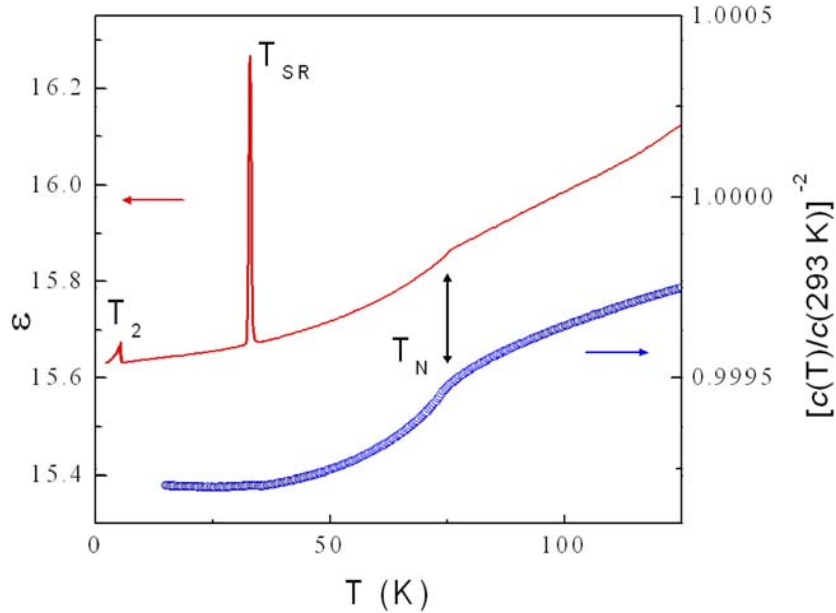

Fig. 8: Comparison of the temperature dependence of the dielectric constant and the inverse square of the c-axis length.

The low-temperature part of the phase diagram of HoMnO$_3$ (Fig. 5b) deserves special attention and the search for structural anomalies caused by the magnetoelastic interaction is of particular interest. We have measured the $c$-axis length as a function of the magnetic field for different constant temperatures. The results shown in Fig. 9 reveal clear anomalies of $c(H)$ at various magnetic phase boundaries. At high temperature (Fig. 9a) the transition across the $T_2$ and $T_1$ phase boundaries are marked by a distinct change of slope of $c(H)$. The magnetoelastic coefficient ($dc/dH$) is enhanced in the INT phase as compared to the HT1 and HT2 phases. This enhancement correlates well with the enhancement of dielectric constant and ac magnetic susceptibility[3] in the intermediate phase. It indicates a softness of the coupled magnetic and lattice systems resulting in the larger response to external electric or magnetic fields. The origin of this softness has to be found in the particular P6'$_3$ magnetic order in the INT phase proposed above. There are two values of $\Phi$ (the angle of the $Mn^{3+}$ spins with the hexagonal $a$-axis) that allow for the higher magnetic symmetry compatible with the space group of the lattice (P6$_3$cm). These values are $\Phi=90º$ (HT1 phase, P6'$_3$c'm) and $\Phi=0º$ (HT2 phase, P6'$_3$cm'). Both magnetic structures appear to be more rigid with respect to external fields. In the INT phase, however, the magnetic symmetry is lower, the angle $\Phi$ adopts values between 0 and 90º and it is more easily affected by external fields resulting in a higher susceptibility. The increase of $c(H)$ in passing from the HT2 phase via the INT phase into the HT1 phase with increasing H is in perfect agreement with the zero-field thermal expansion data that show a sudden increase of $c(T)$ at $T_{SR}$ upon warming.

At the lowest temperatures (T < 3 K) two distinct minima of $c(H)$ are detected between 5 kOe and 20 kOe, as shown in Fig. 9b. These minima reflect the first order phase transition across $T_3(H)$ into the dome shaped LT1 phase. At higher magnetic field (and low T) $c(H)$ increases suddenly at the transition into the LT2 phase and it becomes independent of H in this phase. The magnetoelastic effects observed in different regions of the phase diagram and the structural anomalies at the magnetic phase transitions emphasize the important role of the spin-lattice coupling in $HoMnO_3$ and its correlation with the dielectric anomalies discussed above.

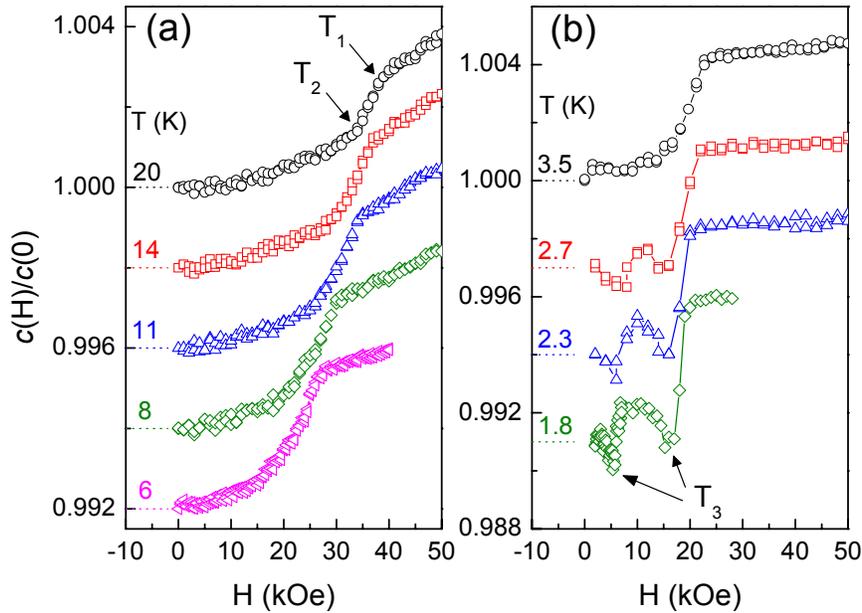

Fig. 9: Magnetostrictive effect on the $c$-axis of $HoMnO_3$. Anomalies of $c(H)$ at the phase boundaries $T_1(H)$, $T_2(H)$, and $T_3(H)$ are clearly identified. Different curves are offset by a constant for better clarity.

SUMMARY

The T-H phase diagram of HoMnO$_3$ is completely resolved via measurements of the temperature and field dependence of dielectric constant for T > 1.8 K and H < 70 kOe. Below the Néel temperature at least six different phases are distinguished through sharp changes of ε(T, H). Two tricritical points are identified in the phase diagram and the possible existence of a tetracritical point close to or at zero temperature is proposed. The complex phase diagram and the observed phase multiplicity requires the additional symmetry breaking such as the P6'$_3$ magnetic symmetry proposed for an intermediate phase that exists between the P6'$_3$c'm and P6'$_3$cm' magnetic phases.

We have provided unambiguous evidence for the existence of extraordinarily strong spin-spin correlations and spin-phonon coupling well above the Néel temperature (as high as room temperature). The negative thermal expansivity of the hexagonal *c*-axis and the thermal expansion anomalies observed at the Néel temperature as well as at the spin reorientation transition are explained by strong magnetic correlations and the magnetoelastic effect. The dielectric anomalies observed at all magnetic transitions, even at the lowest temperatures, are attributed to the spin-lattice coupling.

The complexity of the phase diagram of HoMnO$_3$ and the exciting physical phenomena such as the coupling between magnetic order and dielectric (ferroelectric) properties and their correlations with the lattice degrees of freedom can only by understood if all relevant interactions are taken into account. This includes the Mn$^{3+}$ spins and their AFM superexchange interaction, the Ho$^{3+}$ magnetic moment and their coupling to the Mn spins as well as (at lower temperature) the magnetic exchange between Ho moments, and the ferroelectric polarization including the indirect coupling to the magnetic subsystems via the spin-phonon interaction. This is a difficult problem to solve theoretically. More detailed experimental work needs to be conducted to explore the true nature of the magnetic orders in the various phases (e.g. by neutron scattering experiments) and to understand the physical origin of the observed phenomena.

ACKNOWLEDGEMENTS

This work is supported by NSF Grant No. DMR-9804325, the T.L.L. Temple Foundation, the John J. and Rebecca Moores Endowment, and the State of Texas through the TCSUH at the University of Houston and at Lawrence Berkeley National Laboratory by the Director, Office of Energy Research, Office of Basic Energy Sciences, Division of Materials Sciences of the U.S. Department of Energy under Contract No. DE-AC03-76SF00098.